\documentstyle[12pt]{article}
\begin{document}
\vskip 2 cm
\begin{center}
\Large{\bf QUARKS IN NUCLEI} 
\end{center}
\vskip 1 cm 
\begin{center}
{\bf AFSAR ABBAS} \\
Institute of Physics, Bhubaneswar-751005, India \\
(e-mail : afsar@iopb.res.in)
\end{center}
\vskip 1 cm
\begin{centerline}
{\bf Abstract }
\end{centerline}
\vskip .5 cm

It is generally believed that quarks being confined inside nucleons
should make their presence felt explictly in nuclei only at higher 
energies. However it is shown here that the hole in the centre of 
$ ^{3}H $, $ ^{3}He $ and $ ^{4}He $, the neutron halos in nuclei, 
the $ \alpha - $ and other clustering effects in nuclei and the nuclear 
molecules all basically arise due to the same underlying quark effects. 
Also the role of triton clustering in very neutron rich nuclei is
emphasized. These require the concept of hidden colour states which arise
from the confinement ideas of QCD for the multi-quark systems.

\vskip 2.5 cm

In contrast to the popular belief that nucleonic degrees of freedom are 
all that is required for a consistent description of the ground state 
and low energy properties of nuclei, it is shown here that there are 
specific nuclear effects wherein no way that one can do without invoking 
the quark degrees of freedom. A claim shall be made that only within the 
nucleonic and mesonic degrees of freedom these effects can just not 
be described consistently. This will allow us to make clearcut predictions 
which can help establish the veracity of the claims made herein [1].

The effects that we shall concentrate upon are , the hole at the centre of 
$ ^{3}H $, $ ^{3}He $ and $ ^{4}He $ [2,3],
the neutron halo nuclei [4,5], the clustering effects in nuclei [6,7,8,9] 
and the nuclear molecules [9,10]. 
As we shall see below, this would require an 
understanding of two or more nucleons strongly overlapping over a small
region of $ \leq 1 fm $. This would necessarily imply a study of
multi-quark states in regions $ \leq 1 fm $. This would lead to the concept
of hidden colour [11]. There have been some claims [12] that hidden colour
may not be a useful concept as these coloured states may be rewritten in
terms of asymptotic colour singlet states. Below, we shall show that this
is not always true. In special circumstances, as shall be discussed here,
hidden colour are unambiguously defined states, basic to physics under
consideration.

 As we shall be discussing the structure of nuclei in the ground state, we
should also have the structure of nucleons in the ground state as well.
Hence we view the nucleon as consisting of 3 constituent quarks in the
s-state. Another point that should be borne in mind is that though the 
r.m.s. radius of nucleon is 0.8 fm, it is a very diffuse system with
matter distribution given by
$ \rho (r) = \rho_{0} \exp{ (-mr) } $
and hence matter extends significantly beyond 0.8 fm.

Why does deuteron not have configurations where the two  nucleons overlap
strongly in regions of size $ \leq 1 fm $ to form 6-quark bags ? Why is
deuteron such a big and loose system ? The reason has to do with the
structure of the 6-q bags formed had the two nucleons overlapped strongly.
As per the colour confinement hypothesis the 6-q wave function looks like
[11] :

\begin{equation}
| 6q > = \frac{1}{ \sqrt{5} } | 1 X 1 > + \frac{2}{ \sqrt{5} } | 8 X 8 >
\end{equation}

where '1' represents a 3-quark cluster which is singlet in colour space 
and '8' represents the same as octet in colour space. Hence $ | 8 X 8 > $ 
is overall colour singlet. This part is called the hidden colour 
because as per confinement ideas of QCD
these octets cannot be separated out asymptotically and so manifest
themselves only within the 6-q colour-singlet system. 
This part was found to be  $ 80 \% $
and this would prevent the two nucleons to come together and overlap
strongly [11]. Therefore the hidden colour would manifest itself as
short range repulsion in the region $ \leq 1 fm $ in deuteron. So the two
nucleons though bound, stay considerably away from each other [2,3].

There has been misundersatnding regarding hidden colour in nuclear 
physics. A few scientists have claimed [12] that hidden colour may not be 
a useful concept as these hidden colour states can be rearranged in terms 
of asymptotic colour singlet states. It was pointed out in ref. [12] that 
the hidden colour concept is not unique
only when the two clusters do not overlap strongly and asymptotically can
be separated out. However when the clusters of 3-q each overlap strongly
so that the relative distance between them approacehs zero then the hidden
colour concept becomes relevant and unique as shown in ref. [11].
We would like to point out
that indeed, this necessarily is the situation for deuteron discussed
above. Also note that for the ground state the quark configuration is 
$ s^{6} $ given by configuration space representation [ 6 ] while $
s^{4}p^{2} $ given by [ 42 ] does not come into play as there is not
enough energy to put two quarks into the p-orbital [13].

The author had earlier obtained group theoretically the hidden colour
components in 9- and 12-quark systems [ 2,3]. For the ground state and 
low energy description of nucleons we
assume that $ SU(2)_{F} $ with u- and d-quarks is required. Hence we
assume that 9- and 12-quarks belong to the totally antisymmetric
representation of the group 
$ SU(12) \supset SU(4)_{SF} \otimes SU(3)_{C} $ where
$ SU(3)_{c} $ is the QCD group and  
$ SU(4)_{SF} \supset SU(2)_{F} \otimes SU(2)_{S} $  where S denotes spin.
Note that up to 12-quarks can sit in the s-state in the group SU(12). The
calculation of the hidden colour components for 9- and 12-quark systems
requires the determination of the coefficients of fractional parentage 
for the group $ SU(12) \supset SU(4) \otimes SU(3) $ [ 1,2 ]
which becomes quite complicated for large number of quarks.
The author found that the hidden colour component [ 1,2 ] of the 9-q
system is $ 97.6 \% $ while the 12-q
system is $ 99.8 \% $ ie. is almost completely coloured.

What is the relevance of
these 9- and 12-quark configurations in nuclear
physics ? The A=3,4 nuclei 
$ ^{3}H $, $ ^{3}He $ and $ ^{4}He $ have sizes of 
1.7 fm, 1.88 fm and 1.674 fm respectively. Given the fact that each
nucleon is itself a rather diffuse object, quite clearly in a size
$ \leq 1 fm $ at the centre of these nuclei the 3 or 4 nucleons would
overlap strongly. As the corresponding 9- and 12-q are predominantly
hidden colour, there would be an effective repulsion at the centre keeping
the 3 or 4 nucleons away from the centre. Hence it was predicted by the
author [2,3] that there should be a hole at the centre of 
$ ^{3}H $, $ ^{3}He $ and $ ^{4}He $.
And indeed, this is what is found through electron scattering (see [1,2 ]
for references ). Hence the hole, ie. significant depression in the
central density of 
$ ^{3}H $, $ ^{3}He $ and $ ^{4}He $ is a signature of quarks in this
ground state property.

Neutron halos have been discovered in
several neutron rich nuclei like 
$ ^{6}He, ^{11}Li, ^{11}Be, ^{14}Be $ etc. [4,5]. 
For example the rms radius of 
$ ^{11}Li $ is 3.2 fm while that of $ ^{9}Li $ is 2.3 fm. Hence in this
halo nuclei it is believed that 2n are very loosely bound to a compact
core of $ ^{9}Li $. So also the 2n in 
$ ^{6}He $ etc. It would be fair to say that as of now the halo structur
is an unexplained and puzzling problem in nuclear physics.

 As to $ ^{4}He $ it is a very strongly bound system with binding
energy of 28.29 MeV. It is a compact object of size 1.674 fm. Due to
specific quark hidden colour state, as we discussed above, it has a hole
of size $ \sim 1 fm $ at the centre. Thus it has a significantly higher
density at the boundary and very small at the centre. Hence 
$ ^{4} He $ is like a tennis-ball. Add two more neutrons to 
$ ^{4}He $ to make it $ ^{6}He $, a bound system. This is like adding 2
neutrons to a tennis-ball nucleus. As the two neutrons approach the
surface they will bounce off. As the two neutrons are bound, these
will ricochet on the compact tennis-ball like nucleus. 
A neutron halo would be manifested as these neutrons
shall be kept significantly away from the core.

Macroscopically, as the density of the $ ^{4}He $ core is high on the
boundary, any extra neutrons would not be able to penetrate as this would
entail much larger density on $ ^{4}He $ surface than the system would
allow dynamically. Microscopically, any penetration of extra neutron
through the surface of $ ^{4}He $ would necessarily imply the existence of
five or six nucleons at the centre. As already indicated due to the
relevant SU(12) group only 12-quarks can sit in the s-state, which already
is predominantly hidden colour. Any extra quarks hence would have to go to
the p-orbital and in the ground state of nuclei, there is not sufficient
energy to allow this. Hence the two neutrons are consigned to stay outside
the $ ^{4}He $ boundary. In addition if at any instant the two neutrons
come close to each other while still being close to the surface, locally
the system would be like three nucleons overlapping and looking like a 9-q
system. This too would be prevented by the local hidden colour repulsion.
Hence as found experimentally the two neutrons in the halo would not come
close to each other [4,5].Hence the neutron halo in $ ^{6}He $ is due to
quark effects [1]. About other neutron (and proton) halo nuclei we shall
discuss shortly.

Clusters, especially $ ^{4}He $ ones
have a very important role in nuclear structures [ 5,6,7,8 ]. Clusters are
crucial for studies of light nuclei like 
$ ^{8}Be, ^{7}Be,$
$ ^{6}Li, ^{7}Li, ^{12}C, ^{16}O $. Even for heavier
nuclei like  $ ^{20}Ne, ^{24}Mg, ^{28}Si, ^{44}Ti $ and others they are
important [ 5,6,7,8 ].
It is commonly stated that $ ^{4}He $ clusters are formed in nuclei
because it is so strongly bound, ie. 28.29 MeV. Here we would like to
point out that $^{4}He $ forms good clusters because in addition it has a
hole at the centre so it is like a tennis-ball. These balls in a bound
system of several $ ^{4}He $ nuclei would bounce from each other. 

The reason for resistance to inter-penetration of two $ \alpha $
clusters would be hidden colour repulsion of the relevant
6-, 9- and 12-quark systems
plus the fact that more than 12 quarks are not permitted in the
lowest s- state in the group SU(12).

Let us treat $ ^{12}C $ as $ 3 \alpha $ cluster with the $ \alpha $'s
sitting at the vertices of an equilateral triangle. Because of 
tennis-ball like structure the three $ \alpha $ particles cannot come too
close to each other. Firstly, the surface of the ball would prevent it and
secondly if some part of the 3 $ \alpha $ 's still overlap at the centre,
it would look like a 6- or 9-quark system. Therein the hidden colour
components would repel ensuring that the 3 $ \alpha $ clusters do not
approach too closely at the centre. This too would imply a depression in
the central density of $ ^{12}C $. Indeed, from the density distribution
determination by electron scattering, this is so in $ ^{12}C $.
$ ^{16}O $ treated as $ 4 \alpha $ sitting at the vertices of 
a regular tetrahedron would, for the
reasons stated above, too have a central density depression, again as seen
in the electron scattering. Due to the central depression, 
$ C^{12} $ and $ ^{16}O $ would appear tennis-ball like as well [1].

In our picture $ ^{16}O $ of 4 $ \alpha $ at the vertices of a regular 
tetrahedron is special. Just as $ ^{4}He $ with
12-q structure at centre is special due to the degeneracies of the group
SU(12) so also $ ^{16}O $ with 4 $ \alpha $
is special for the same reason. For $ ^{20}Ne $ the fifth $ \alpha $
would not just penetrate $ ^{16}O $ but would try to form 
a regular tetrahedron locally with any of the 3 $ \alpha $ surface
of the  4 $ \alpha $ cluster. Similarly we can keep on adding $ \alpha $
's until $ ^{32}S $ where 4 $ \alpha $
's are sitting on top of four 3 $ \alpha $
clusters surfaces of $ ^{16}O $. 
Hence for $ \alpha $'s 5 to 8 it is close packing on top of 4 $ \alpha $'s
of $ ^{16}O $ which has a hole at the centre. Hence all these nuclei
should have hole at the center.
Beyond 8 $ \alpha $ clusters, the geometry is such that the 9th $ \alpha $
- particle cannot be simply close packed on top of $ ^{32}S $.

It has been shown here that the $ ^{12}C $ with 3 $ \alpha $
and $ ^{16}O $ with 4 $ \alpha $
has central density depression. Though the effect may be softened compared
to $ ^{3}H $, $ ^{3}He $ as 3N and and $ ^{4}He $ as 4N ball structures,
nevertheless these nuclei should also act tennis-ball like. No wonder one
observes nuclear molecules in C-C, C-O,
O-O systems. The stability of nuclear molecules has to do with the
bounciness of the C,O nuclei coupled with the fact that their large size
is due to the largeness of the constituent clusters. 
Also obviously molecular structures would exist  for nuclei $ ^{20}Ne,
^{24}Mg, ^{U28}Si $ and $ ^{32}S $ [9,10]. Hence nuclear
molecules [9,10] too arise basically due to quark effects [1].

So far we have explained neutron halo nuclei $ ^{6}He $ as 
arising due to 2n ricochet off the stiff tennis-ball like core of
$ ^{4}He $. In addition, nuclei like 
$ ^{12}C $ are made up of 3 ball like 
$ \alpha $'s and also develops ball like properties.
What happens when 2n are added to it ? Could one have two neutron halos
for $ ^{14}C $ ? This is not so. The reason is because of the following.
  Going through the binding energy systematics of neutron rich nuclei one
notices that as the number of 
$ \alpha $'s increases along with the neutrons, each $ ^{4}He $ + 2n pair
tends to behave like a cluster of two 
$ ^{3}_{1}H_{2} $ nuclei. Remember that though 
$ ^{3}_{1}H_{2} $ is somewhat less strongly bound (ie. 8.48 MeV ) it is
still very compact (ie. 1.7 fm ), almost as compact as $ ^{4}He $ (1.674
fm). In addition it too has a hole at the centre. Hence $ ^{3}H $ is also
tennis-ball like nucleus. This splitting tendency of neutron rich nuclei
becomes more marked as there are fewer and fewer of $ ^{4}He $ nuclei left
intact by the addition of 2n. Hence $ ^{7}Li $ which is 
$ ^{4}He + ^{3}H $ with 2n becomes $ ^{9}Li $ which can be treated as made
up of $ 3 ~^{3}H $ clusters and should have hole at the centre. Similarly
$ ^{12}Be $ consists of 4 $ ^{3}_{1}H_{2} $ 
clusters and $ ^{15}B $ of 5 $ ^{3}_{1}H_{2} $  clusters etc.
Other evidences like the actual decrease of radius as one goes from
$ ^{11}Be $ to $ ^{12}Be $ supports the view that it ( ie $ ^{12}Be $ )
must be made up of four compact clusters of $ ^{3}H $.

Note that just as several light N=Z nuclei with A=4n, n=1,2,3,4 ... can be 
treated as made up of n  $ \alpha $ clusters, in Table 1 we show several 
neutron rich nuclei which can be treated as made up of n
$ ^{3}_{1}H_{2} $ clusters. We can write the binding energy of these
nuclei as [1]

\begin{equation}
E_{b} = 8.48 n + Cm 
\end{equation}
 
Here  n 
$ ^{3}_{1}H_{2} $ clusters form m bonds and where C is the inter-triton
bond energy. We have assumed  the same geometric structure of clusters in
these nuclei as for $ \alpha $ clusters of A = 4n nuclei as given above.
All the bond numbers arise due to
these configurations. We notice from Table that this holds good and 
and that the inter-triton cluster bond energy is approximately 5.4 MeV.
This value seems to work for even heavier neutron rich
nuclei. For example for 
$ ^{42}Si $ the inter-triton cluster energy is still 5.4 MeV. Notice that
the geometry of these cluster structures of $ ^{3}H $ becomes more complex
as the number increases but nevertheless still holds good.

What we are saying is that
these neutron rich nuclei, made up of n number of
tritons, each of which is tennis-ball like and compact, should be compact
as well. These too would develop tennis-ball like property. This is
because the surface is itself made up of tennis-ball like clusters. 
Hence as there are no more $ ^{4}He $ clusters to break when
more neutrons are added to this ball of triton clusters, these 
extra neutrons will ricochet on the surface.
Hence we expect that one or two neutrons outside these compact clusters
would behave like neutron halos. Therefore $ ^{11}Li $ with $^{9}Li + 2n $
should be two neutron halo nuclei - which it is [ 3,4 ]. So should 
$ ^{14}Be $ be [ 3,4 ]. It turns out that internal dynamics of $ ^{11}Be $
is such that it is a cluster of $ \alpha - t - t $ ( which also has to do
with $ ^{9}Li $ having a good 3 $ \alpha $ cluster)
with one extra neutron halo around it. Next $ ^{17}B, ^{19}C, ^{20}C $
would be neutron halo nuclei and so on [1]. These specific predictions
of quark signatures in nuclei are eaily testable.

Thus all light neutron rich nuclei $ _{Z}^{3Z}A_{2Z} $ are made up of Z 
$ ^{3}_{1}H_{2} $ clusters. Due to hidden colour considerations
arising from quark effects, all these should have holes at the centre.
This would lead to tennis-ball like property of these nuclei.
One or two (or more) extra neutrons added to these core nuclei 
would ricochet on the surface of the core nucleus and
form halos around it. All known and well-studied neutron halo
nuclei fit into this pattern. This
makes unambiguous predictions about which nuclei should be neutron halo
nuclei and for what reason. At the base, it is quarks which cause neutron
halo [ 3,4 ]. The proton halo nuclei can also be understood in the same
manner. Here another nucleus with a hole at the centre 
$ ^{3}_{2}He_{1} $ (binding energy 7.7 MeV, size 1.88 fm) would play a
significant role.

So it should be clear now that
it is quarks through hidden colour configuration lead
to a hole at the centre of 
$ ^{3}H $, $ ^{3}He $ and $ ^{4}He $. As the relevant group is SU(12) no
more than 12 quarks can sit in the lowest orbital. Hence the hole at the
centre of $ ^{4}He $ is special. Clustering of these nuclei is also
determined by their bouncing tennis-ball like property.
This gives new insight into $ \alpha - $ clustering in nuclei and predicts
existence of clusters of triton ( and helion ) nuclei. All these nuclei
are themselves compact and tennis-ball like. One, two or more neutrons 
outside these nuclei ricochet to give halo like structures. Large nuclear 
molecules can also be understood in the same manner. 
Hence these should be treatd as unique quark signatures in nuclei.

\newpage

\vskip 7.0 cm

\begin{table}
\centerline {\bf TABLE } 
\centerline{\bf Neutron rich nuclei with inter-triton cluster 
bond energies}
\vskip 0.2 in
\begin{center}
\begin{tabular}{|c|c|c|c|c|}
\hline
Nucleus & n & m & $ E_{B} - 8.48n (MeV) $ & C(MeV) \\
\hline
$ ^{9}Li $ &  3 & 3 & 19.90 & 6.63 \\
\hline
$ ^{12}Be $ & 4 & 6 & 34.73 & 5.79   \\ 
\hline
$ ^{15}B $ & 5 & 9 & 45.79 & 5.09 \\
\hline
$ ^{18}C $ & 6 & 12 & 64.78 & 5.40 \\
\hline
$ ^{21}N $ & 7 & 15 & 79.43 & 5.29 \\
\hline
$ ^{24}O $ & 8 & 18 & 100.64 & 5.59 \\
\hline

\end{tabular}
\end{center}
\end{table}

\newpage

\begin{center}
{\bf REFERENCES }
\end{center}

1. A. Abbas, {\it Mod. Phys. Lett.} {\bf A 16} (2001) 755

2. A. Abbas, {\it Phys.Letts.} {\bf 167 B} (1986) 150

3. A. Abbas, {\it Prog. Part. Nucl. Phys.} {\bf 20} (1988) 181 

4. I. Tanihata, {\it J. Phys.} {\bf G22} (1996) 157 

5. P. G. Hansen, {\it Nucl. Phys.} {\bf A 553} (1993) 89c

6. P. E. Hodgson, {\it Z. Phys.} {\bf A 349 } (1994) 197

7. A. C. Merchant and W. D. M. Rae, {\it Z. Phys. } {\bf A 349 } (1994)
243

8. J. S. Lilley and M. A. Nagarajan, Eds, 
{\it `Clustering Aspects of Nuclear Structure,' } D. Reidel Publ. Co.,
Dodrecht, Holland, 1985.

9. B. R. Fulton, {\it Z. Phys. } {\bf C 349 } (1994) 227

10. W. Greiner, J. Y. Park and W. Scheid, 
{\it `Nuclear Molecules' }, World Scientific Publishing Co. Ltd.,
Singapore, 1995.

11. V. A. Matveev and P. Sorba, 
{\it Lett. al Nuovo Cim.} {\bf 20} (1977) 435

12. P. Gonzalez and V. Vento, 
{\it Il Nuovo Cimento} {\bf 106 A} (1992) 795

13. M. Harvey, {\it Nucl. Phys.} {\bf A 352} (1981) 301, 326

\end{document}